\documentclass[twocolumn,showpacs,preprintnumbers,amsmath,amssymb,prb]{revtex4}

\usepackage{graphicx}
\usepackage{dcolumn}
\usepackage{bm}

\begin{document}


\title{Phase coherent transport in a side-gated mesoscopic graphite wire}

\author{D. Graf}
\email{grafdavy@phys.ethz.ch}
\author{F. Molitor}
\author{T. Ihn}
\author{K. Ensslin}
\affiliation{Solid State Physics Laboratory - ETH Zurich, Switzerland}

\date{\today}

\begin{abstract}
We investigate the magnetotransport properties of a thin graphite wire resting on a silicon oxide substrate. The electric field effect is demonstrated with back and side gate electrodes. We study the conductance fluctuations as a function of gate voltage, magnetic field and temperature. 
The phase coherence length extracted from weak localization is larger than the wire width even at the lowest carrier densities making the system effectively one-dimensional. We find that the phase coherence length increases linearly with the conductivity suggesting that at 1.7 K dephasing originates mainly from electron-electron interactions. 
\end{abstract}

\maketitle


\section{Introduction}
Graphite is composed of stacked layers of graphene sheets in which carbon atoms are ordered in a two-dimensional hexagonal lattice. It is a semimetal with equal electron and hole densities in the undoped case. Recently, it has been shown that thin graphite flakes of a few nanometers in height exhibit a pronounced electric field effect.\cite{Zhang04, Novoselov04, Zhang05} The applied potential is screened on a length scale corresponding to the interlayer distance implying that the back gate electrode only affects the first few graphene layers close to the insulating substrate. A single-layer of graphene ultimately confines the carriers in a sheet of atomic thickness: the electronic bandstructure is, however, modified resembling a gapless semiconductor with a linear energy dispersion relation. Well-defined plateaus were measured in the quantum Hall effect opening the way to investigate properties observed so far to two-dimensional electron and hole gases at the interfaces of layered semiconductors.\cite{Novoselov05b, Zhang05b}

We report low-temperature magnetotransport measurements on a few-layer graphene wire whose conductance is tunable both with back and side gate electrodes. The combined observation of weak localization and magnetoconductance fluctuations shows that the system is mesoscopic, one-dimensional and in the diffusive regime. The extracted phase coherence length varies from 0.5 $\mu$m up to 2.5 $\mu$m for estimated carrier densities from zero to 2.5$\times$10$^{12}$/cm$^2$. While this regime has been extensively studied in GaAs/AlGaAs systems, \cite{Beenakker91} very few experiments have been reported for single- and few-layer graphene.\cite{Morozov05, Berger06} We find the phase coherence length to be proportional to the conductivity suggesting that the main dephasing mechanism at low temperatures is related to electron-electron collisions with small energy transfer.\cite{Altshuler82} Finally, adding the electric field effect contributions of the back and side gates allows us to vary the disorder configuration at a given Fermi level. 

\begin{figure}
\includegraphics[width=0.4\textwidth]{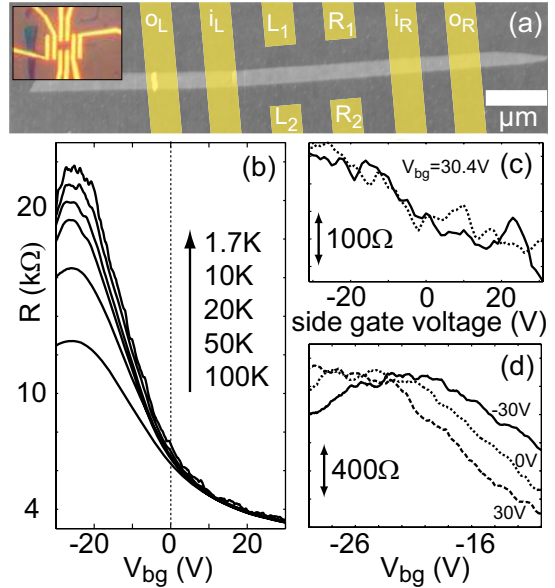}
\caption{\label{fig:sample} (a) SFM micrograph of a graphite wire resting on a silicon oxide surface with a schematic of the four contacts (i$_\mathrm{L}$, i$_\mathrm{R}$, o$_\mathrm{L}$, o$_\mathrm{R}$) and four side gates (L$_1$, L$_2$, R$_1$, R$_2$). Inset: Optical microscope image of the structure. (b) Four-terminal resistance as a function of back gate for different temperatures. (c) Resistance change as a function of the side gates L$_1$+L$_2$ (solid line) and R$_1$+R$_2$ (dotted line) at 1.7 K. (d) Resistance change as a function of back gate for different side gate voltages (L$_1$+L$_2$+R$_1$+R$_2$) at 1.7 K.}
\end{figure}

\section{Sample and setup}
Deposition of graphite (highly oriented pyrolitic graphite, HOPG) by mechanical exfoliation produces flakes with a large variety of shapes,\cite{Novoselov04} among them also wires, sub-micrometer in width but several micrometers in length. The wire investigated in this paper is shown in Fig. \ref{fig:sample}(a). It has a width $W$=320 nm and is 3.2 nm high corresponding to 7 $\pm$ 1 stacked layers of graphene. 
The thickness is determined with a scanning force microscope. Alternatively information from the Raman spectrum can be used to count the number of layers for flakes with a larger lateral extent.\cite{Graf07} 
Cr(5 nm)/Au (90 nm) contacts and side gates (with width and gaps of 0.5 $\mu$m) are evaporated onto and next to the wire (Fig. \ref{fig:sample}(a) inset). The wire length measured between the centers of the two inner contacts (i$_\mathrm{L}$,i$_\mathrm{R}$) is $L$=3 $\mu$m. By extrapolating the two-terminal resistances between pairs of contacts to zero distance we find a contact resistance of $R_{\mathrm{contact}} \approx$ 2 k$\Omega$, while the direct comparison between the two- and the four-terminal configuration results in $R_{\mathrm{contact}} \approx$ 1 k$\Omega$. Subsequent measurements are all done in a four-point setup by applying current (ac, 100 nA rms) through the outer contacts (o$_\mathrm{L}$,o$_\mathrm{R}$) and measuring the voltage difference between the inner contacts (i$_\mathrm{L}$,i$_\mathrm{R}$, labeled in Fig. \ref{fig:sample}(a)). 

\section{Results and discussion}
\subsection{Electric field effect}\label{sec:efe}
In case of graphite the screening length is of the order of the interlayer distance: $\lambda_\mathrm{s} \approx$ 0.4 nm.\cite{Visscher71} Making graphite samples thinner than approximately 50 nm will eventually lead to a measurable field effect since the proportion of the induced charge density to the unaffected bulk charge density becomes significant.\cite{Zhang04,Novoselov04} Resistance traces as a function of back gate voltage are shown in Fig. \ref{fig:sample}(b) for temperatures from 1.7 K up to 100 K. Two distinct regimes can clearly be identified (see dotted line in Fig. \ref{fig:sample}(b)): Around the resistance maximum ($V_{\mathrm{bg}}\approx$ -24 V) we find a pronounced decrease in resistance with increasing temperature, whereas for large positive back gate voltages the resistance is almost independent of temperature.

These two findings can be explained within the simple two band (STB) model (see, e.g., Ref. [\onlinecite{Klein64}]), where the bandstructure of graphite is represented by overlapping parabolic valence band ($E=E_0/2-\hbar^2k^2/2m^*$) and conduction band ($E=-E_0/2+\hbar^2k^2/2m^*$) dispersions. 
The three-dimensional density of states for valence and conduction band is taken to be $4 m^*/\pi \hbar^2 c_0$, where $c_0$ is twice the interlayer distance. Taking into account an exponential decay of the applied potential we find for the lowest temperature ($T$=1.7 K) an energy overlap $E_0$=2.8 meV and an effective electron mass of ${m_\mathrm{e}}^*$=0.041$m_\mathrm{e}$, which agrees well with previously reported data on thin graphite flakes.\cite{Zhang05,Novoselov04,Morozov05} Near the resistance maximum the Fermi-Dirac distribution will at finite temperature populate more electron and hole states at the band edges compared to the sharp energy cutoff in the zero temperature limit leading to an enhanced carrier density and thus to a reduced sample resistance [$R=L/W(n_\mathrm{e}+n_\mathrm{p})e\mu$]. For large back gate voltages far away from the mixed region in the regime of pure electron transport the smearing of the Fermi edge will not change the overall density as long as $k_\mathrm{B} T \ll E_\mathrm{F}$. 
The electron mobility estimated in this regime by combining a simple parallel plate capacitor model for the induced electron density and the quasi-linear increase in conductivity yields about 3200 cm$^{2}$/Vs at $T$=1.7 K compared to the 2700 cm$^{2}$/Vs extracted from the above mentioned model which includes also the mixed region. It is two orders of magnitude smaller than for macroscopic samples of HOPG at low temperatures\cite{Morozov06} suggesting that the mean free path $l_{\mathrm{e}} \approx$ 70 nm is smaller than the wire width.\cite{Novoselov04} 

\begin{figure}
\includegraphics[width=0.4\textwidth]{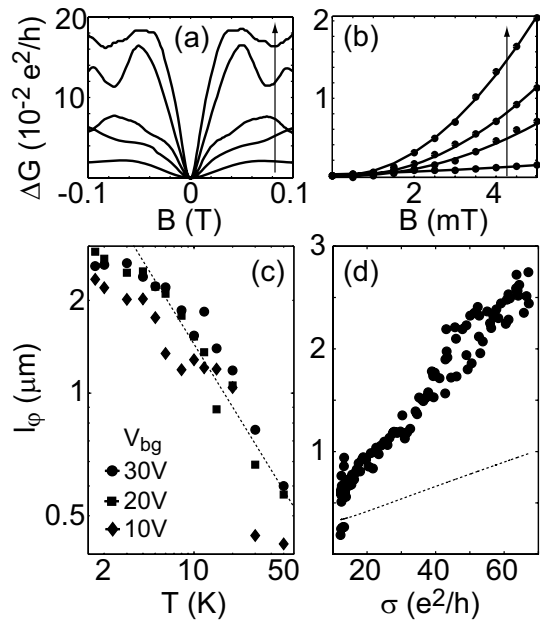}
\caption{\label{fig:wl} Weak localization (a) for increasing back gate voltage (direction of the arrow: -10, 0, 10, 20, 30 V) and (b) for increasing temperature (dots, direction of the arrow: 30, 12, 6, 2 K) with corresponding fits using Eq. \ref{eq:wl1D} (solid lines). All the curves are normalized to $B$=0 T. (c) Phase coherence length as a function of temperature for back gate voltages V$_{\mathrm{bg}}$=10, 20, 30 V in a double logarithmic plot at 1.7 K. Dotted line represents a power law fit in the range $T$ $\ge$ 4 K. (d) Phase coherence length as function of conductivity. Dots: Measured data. Dotted line: Altshuler-Aronov-Khmelnitsky dephasing.}
\end{figure}

Additional electrical control is gained via side gates. Applying voltages just to two opposite gate fingers and leaving the other two grounded we can compare the lateral field effect in two spatially separate segments of the graphite wire (see Fig. \ref{fig:sample}(c)). The resistance traces for voltages applied either to L$_1$+L$_2$ (solid line) and to R$_1$+R$_2$ (dotted line) have the same slope even though they differ in the details of the superimposed reproducible fluctuations. Using all four side gates with voltages of $\pm$30 V shifts the whole resistance curve as a function of back gate voltage by $\pm$ 3.1 V as shown in Fig. \ref{fig:sample}(d), but does not change the shape of the curve qualitatively, except for the details of the superimposed reproducible fluctuations. The lever arm of all side gates is thus ten times smaller than that of the back gate changing the Fermi energy only within the mixed region ($\Delta E_{\mathrm{F}} \approx E_0$).

\begin{figure}
\includegraphics[width=0.4\textwidth]{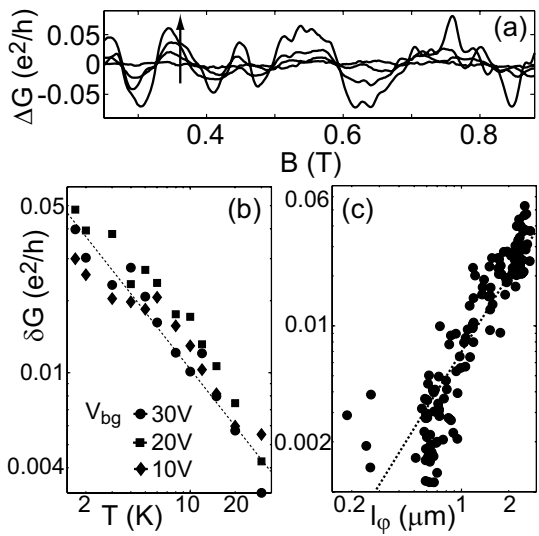}
\caption{\label{fig:ucfB} (a) Magnetoconductance fluctuations for different temperatures (30, 10, 5, 1.7 K) at fixed back gate $V_{\mathrm{bg}}$=30 V and grounded side gates. (b) Conductance fluctuation amplitude extracted from the magnetic field range shown in (a) plotted versus temperature for $V_{\mathrm{bg}}$=10, 20, 30 V (dot, square, diamond). The dotted line follows a power law in $T$. (c) Conductance fluctuation amplitude (dots) in the magnetic field range shown in (a) plotted as function of the phase coherence length (taken from Fig. \ref{fig:wl}(d)) at $T$=1.7 K. Dotted line corresponds to a power law fit.}
\end{figure}

\subsection{Weak localization}
In samples with a phase coherence length larger than the elastic scattering length, quantum corrections of the Drude resistance due to constructive interference of time-reversed paths lead to an enhanced resistance at zero magnetic field. Figs. \ref{fig:wl}(a) and (b) show typical conductance traces at low magnetic field exhibiting this weak localization (WL) effect for increasing back gate voltage (-10, 0, 10, 20, 30 V) and for decreasing temperature (30, 12, 6, 2 K). For low temperatures and high back gate voltage the curvature is more pronounced and the peak amplitude larger.
From the suppression of the weak localization in a perpendicular magnetic field we extract the phase coherence length by fitting the one-dimensional expression for $l_\mathrm{e}$ $\ll$ W $\ll$ $l_{\phi}$ (dirty metal regime):\cite{Beenakker91}
\begin{eqnarray}\label{eq:wl1D}
	\delta G_{\mathrm{loc}}^{\mathrm{1D}}(B)=-\frac{2e^2}{h}\frac{1}{L}\left(\frac{1}{D\tau_{\phi}}+\frac{1}{D\tau_{B}}\right)^{\frac{1}{2}}
\end{eqnarray}
where $\tau_{B}\approx {l_\mathrm{m}}^4/(DW^2)$, and $l_\mathrm{m}=(\hbar/eB)^{1/2}$. Here W is the lithographic width. When $W \ge l_\mathrm{m} (B)$ the 2D formalism must be used, since the lateral extension becomes irrelevant,\cite{Beenakker91} which in our case  restricts the fit to a field range below 6.5 mT.

In Fig. \ref{fig:wl}(c) the phase coherence length l$_\phi=\sqrt{D\tau_{\phi}}$ extracted using Eq. (\ref{eq:wl1D}) is plotted as a function of temperature for three different back gate voltages. We find an exponent of $\nu \approx -0.69$ by analyzing the temperature dependence ($T$ $\ge$ 4 K) and assuming l$_{\phi} \propto T^{\nu}$. In Ref. [\onlinecite{Berger06}] a power law $l_{\phi} \propto T^{-2/3}$ is used to fit the WL data of a wire cut out of ultrathin epitaxial graphite. In Ref. [\onlinecite{Morozov06}] $l_{\phi}$ was found to vary approximately as $T^{-1/2}$ for single-layer graphene.
The maximum phase coherence length in our experiment for $T$ $<$ 4 K lies between 2.5 and 3 $\mu$m and suggests that transport between the two inner contacts is fully phase coherent. The saturation of the phase coherence length at low temperatures is thus attributed to the extension of the coherent region into the leads.

In Fig. \ref{fig:wl}(d) the phase coherence length at 1.7 K is found to be linear as function of conductivity ($\sigma=G \times L/W$). In the experiment the back gate voltage is changed in order to tune the conductivity. At these low temperatures dephasing is usually attributed mainly to electron-electron interactions.\cite{Altshuler82} The dephasing rate in the two-dimensional case ($l_e \gg$ $W$) is given by \cite{Chakra86}
\begin{eqnarray}\label{eq:altsh}
	\frac{\hbar}{\tau_{\phi}}=kT\frac{e^2/\hbar}{\sigma}\mathrm{ln}\left(\frac{kT}{\hbar/\tau_{\phi}}\right) \approx kT \frac{e^2/\hbar}{\sigma} \mathrm{ln}\left(\frac{\sigma}{e^2/\hbar}\right)
\end{eqnarray}
where the approximation holds for $\sigma \gg e^2/\hbar$. Extracting the decoherence time from the above equation and using the Einstein relation $\sigma=e^2 D {\cal{D}}_{\mathrm{2D}}$ with the 2D density of state ${\cal{D}}_{\mathrm{2D}}$ and the diffusion constant $D$, we find that $l_{\phi}$ is linear in $\sigma$ with a logarithmic correction. With the above extracted effective mass for electrons ${m_\mathrm{e}}^*$=0.041$m_\mathrm{e}$ the theoretical prediction follows the dotted line in Fig. \ref{fig:wl}(d). Similar discrepancies have been found for 1D wires on AlGaAs heterostructures.\cite{Senz}

\subsection{Conductance fluctuations}
In mesoscopic physics the disorder configuration matters when the size of the conductor is of the order of the phase coherence length. As a result magnetoconductance fluctuations are superimposed on the classical Drude conductance.\cite{Beenakker91} The fluctuations are reproducible and parametric in the electric or magnetic field.

In Fig. \ref{fig:ucfB}(a) conductance (with a linear background subtracted) as a function of magnetic field is shown for decreasing temperature for a fixed back and side gate setting. In this low field regime Landau quantization can be neglected. The fluctuations can be continuously traced from 30 K down to 1.7 K. Between successive curves the back gate was swept from -31 to 31 V. We conclude that the static disorder is stable and thus characteristic for a given cool down. 

A measure for the strength of the fluctuations is the root-mean-square magnitude of the conductance fluctuations $\delta G_{V_{\mathrm{bg}}/\mathrm{B}}$ (from now on referred to as conductance fluctuation amplitude).
In Fig. \ref{fig:ucfB}(b) the temperature dependence of $\delta G_{\mathrm{B}}$ extracted for the magnetic field interval shown in Fig. \ref{fig:ucfB}(a) is presented in a double logarithmic plot for three positive back gate voltages. Even for the lowest temperatures the conductance fluctuation amplitude does not saturate but follows a power law $T^{\nu}$ with an exponent estimated to $\nu \approx$ -0.8 indicated by the dotted line.\cite{Lee85} 

In Fig. \ref{fig:ucfB}(c) the conductance fluctuation amplitude is plotted versus the phase coherence length extraced from weak localization. For a narrow channel with $l_{\phi}$ larger than the wire width W and smaller than the wire length $L$ a power law
\begin{eqnarray}\label{eq:narrow}
\delta G \propto \frac{e^2}{h}\left(\frac{l_{\phi}}{L}\right)^{\nu}
\end{eqnarray}
links both quantities.\cite{Beenakker91} The dotted line corresponds to a fit to the above equation: the extracted exponent is $\nu$ = 1.64, which is in good agreement with the theoretically expected value of $\nu$=1.5.

\begin{figure}
\includegraphics[width=0.4\textwidth]{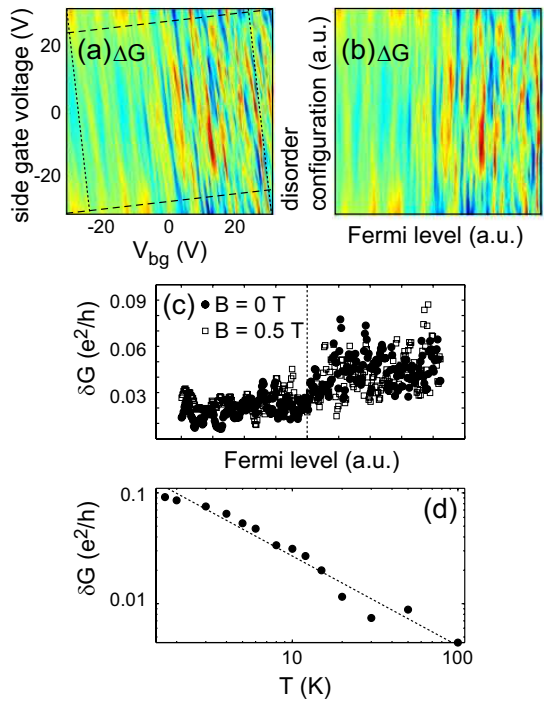}
\caption{\label{fig:ucfgate} (a) Conductance fluctuations as a function of back gate and side gate voltage. A linear background for constant back gate voltage has been substracted from the data (see linear increase in Fig. \ref{fig:sample}(c)). (b) Conductance fluctuations as a function of Fermi level (dashed line in (a)) and disorder configuration (dotted line in (a)). (c) Conductance fluctuation amplitude along the disorder configuration line as a function of the Fermi level for $B$=0 T (dots) and $B$=0.5 T (square, scaled with 1/$\sqrt{2}$). (d) Conductance fluctuation amplitude (dots) in the back gate range of 10-30 V as a function of temperature along with a power law fit (dotted line).}
\end{figure}

With the additional side gates, conductance fluctuations at constant Fermi energy and carrier density can be studied. In Fig. \ref{fig:ucfgate}(a) the conductance is plotted as a function of back and side gates voltages. In Fig. \ref{fig:ucfgate}(b) the coordinate system has been rotated with respect to (a) taking into account the relative lever arms such that the new horizontal axis corresponds to a change in Fermi energy (mainly caused by the back gate over a range of $\pm$ 31 V) whereas the vertical axis follows changes in the disorder configuration (mainly caused by the four side gates in a voltage range between -31 and 31 V). The STB model introduced in Sec. \ref{sec:efe} yields for the Fermi level (from left to right) a linear increase from $E_\mathrm{F} $=-7.7 meV to 108 meV corresponding to a change in the induced electron and hole density of $n$=2$\times$10$^{11}$/cm$^{2}$ passing zero to 2.3$\times$10$^{12}$/cm$^{2}$. 
Compared to earlier investigations in a GaAs/AlGaAs quantum wire,\cite{Heinzel00} the lack of lateral quantization does not allow for ballistic modes to propagate. Nevertheless we find conductance features mainly along constant density lines. This can be understood as a fixed disorder configuration monitored by the combined back and side gate action.
The conductance fluctuation amplitude calculated along the disorder configuration for fixed Fermi level, Fig. \ref{fig:ucfgate}(c), can be divided into two regions congruent with the one discussed in conjunction with Fig. \ref{fig:sample}(b) (see dotted line): Low amplitude fluctuations for the Fermi energy lying in the region of overlapping valence and conduction band ($k_\mathrm{F} l_\mathrm{e} \approx 1$) and larger amplitude fluctuations for pure electron transport ($k_\mathrm{F} l_\mathrm{e} \gg 1$). This is analogous to the strength of the WL signal shown in Fig. \ref{fig:wl}(b), where the peak amplitude increases for larger positive back gate voltages and thus higher electron densities.
Note that the conductance fluctuations are never completely suppressed. Since the conductance fluctuations are limited to the range where $l_\mathrm{e} < l_{\phi}$ we can infer that the mean free path does not exceed the minimum phase coherence length of about 0.5 $\mu$m shown in Fig. \ref{fig:wl}(d) confirming the estimation drawn from the mobility in Sec. \ref{sec:efe}.

For finite magnetic field time-reversal symmetry breaks down. The conductance fluctuation amplitude scales as $\delta G = \sqrt{Var(G)} \propto \beta^{-1/2} e^2/h$ where $\beta=1$ for zero and $\beta=2$ for finite magnetic field.\cite{Beenakker91} In Fig. \ref{fig:ucfgate}(c) the data for B=0.5 T (square) is corrected by this factor and collapses onto the data collected at zero magnetic field.

In Fig. \ref{fig:ucfgate}(d) the conductance fluctuation amplitude $\delta G_{V_{\mathrm{bg}}}$ determined in the back gate range from 10 to 30 V is shown as a function of temperature and is found to decrease as a power law as pointed out in conjunction with Fig. \ref{fig:ucfB}(b). The fitted slope of -0.8 in the double logarithmic plot is the same as in the case of the magnetic field induced conductance fluctuations.

\section{Conclusion}
In summary, we have shown magnetotransport measurements on an ultrathin graphitic wire and find several properties characteristic for mesoscopic samples in the diffusive regime. The phase coherence length exceeds the wire width and is comparable to the wire length at low temperatures and high electron densities resulting in a fully coherent one-dimensional conductor. The conductance fluctuation amplitudes follow power laws in temperature as well as in phase coherence length.
The proportionality of the conductivity and the phase coherence length indicate that dephasing happens through electron-electron interaction at low temperature. Side gate fingers in addition to the standard back gate electrode allow us to tune disorder and carrier density independently. 

\begin{acknowledgments}
We acknowledge stimulating discussions with K.S. Novoselov and R. Leturcq.
Financial support from the Swiss Science Foundation (Schweizerischer Nationalfonds) is gratefully acknowledged. 
\end{acknowledgments}


\end{document}